\documentclass[a4paper]{jpconf}
\usepackage{graphicx}
\usepackage{amssymb, amsmath, mathbbol}
\begin{document}
\title{Theoretical aspects of charged Lepton Flavour Violation }

\author{Ana M. Teixeira}

\address{Laboratoire de Physique Corpusculaire, CNRS/IN2P3 -- UMR 6533,\\
Campus Universitaire des C\'ezeaux, 4 Av. Blaise Pascal,
63178 Aubi\`ere Cedex, France}

\ead{ana.teixeira@clermont.in2p3.fr}

\begin{abstract}
If observed, charged lepton flavour violation  is a clear sign
of new physics - beyond the Standard Model minimally extended to accommodate
neutrino oscillation data. 
We briefly review several extensions of the Standard Model which could potentially
give rise to observable signals, also emphasising the r\^ole of
charged lepton flavour violation
in probing such new physics models.\footnote{PCCF RI 16-04}   
\end{abstract}

\section{Introduction}
Of the three observations which cannot be explained by the
Standard Model (SM) - and which also include the baryon asymmetry of the
Universe (BAU) and dark matter (DM) - 
neutrino oscillations provided the first evidence of new
physics. Interestingly, among the several models successfully accounting 
for and explaining $\nu$-data, many offer the possibility to further
address the BAU via leptogenesis, succeed in putting 
forward viable DM candidates, or
even ease certain theoretical puzzles of the SM, as is the case of the
flavour problem. 

Leptonic mixings and massive neutrinos offer a true gateway
to many new experimental signals or deviations from SM predictions in
the lepton sector; among others, these include charged lepton
flavour violation (cLFV). 
The most minimal extension of the SM allowing to accommodate $\nu$
oscillation data consists in the addition of right-handed neutrinos
($\nu_R$) while preserving the total lepton number, 
thus giving rise to massive Dirac neutral leptons. In such a
framework, individual lepton numbers are
violated  (as encoded in the $U_{\rm{PMNS}}$ matrix), and 
cLFV transitions such as $\mu \to e \gamma$ can occur, being
mediated by $W^\pm$ bosons and 
massive neutrinos; however, and due to the tiny values of light neutrino
masses, the associated rate is extremely small,
BR($\mu \to e \gamma $)$ \sim \mathcal{O}(10^{-55})$, lying beyond the
reach of any future experiment. Thus, the observation of a cLFV process
would necessarily imply the existence of new physics degrees of
freedom (beyond minimal extensions via massive Dirac neutrinos). 
A comprehensive review of the experimental status of a vast array of
cLFV observables was presented at this Conference 
by W. Ootani~\cite{Ootani.Neutrino16}.

Whether or not charged and neutral LFV are related, or equivalently
if cLFV arises from the mechanism of $\nu$-mass generation, and
the question of 
which is the New Physics model that can be at the origin of these phenomena, 
have constituted the starting point to extensive studies. 
Here we very briefly review the prospects for cLFV observables of 
some appealing and well motivated SM extensions.

\section{cLFV and New Physics models}
Interpreting experimental data on cLFV - be it in the form of a possible
measurement or improved bounds - requires an underlying theoretical
framework: new physics models can lead to ``observable'' cLFV
introducing new sources of lepton flavour violation, as well as new operators
at the origin of the flavour violating transitions and decays. 

A first, model-independent approach consists in parametrising cLFV
interactions by means of higher-order non-renormalisable (dimension $d>4$)
operators. The new low-energy effective Lagrangian can be
written as 
$\mathcal{L}^\text{eff} = \mathcal{L}^\text{SM} + \sum_{n \geq 1}
\mathcal{C}_{ij}^{4+n} \, \Lambda^{-n} \, \mathcal{O}_{ij}^{4+n}$, in
which $\Lambda$ denotes the scale of new physics, and $\mathcal{C}$, 
$\mathcal{O}$ the effective couplings and operators, with the former
corresponding to complex matrices in flavour space. Contrary to the
unique dimension-five Weinberg operator (common to all models with
Majorana neutrino masses), there exists a large number of dimension-six
operators, whose low-energy effects include cLFV. 
Regarding the cLFV dimension-six operators, 
these can be loosely classified as dipole, four-fermion and
scalar/vector operators. 

In order to constrain the new physics scale
and the amount of flavour violation thus introduced, the cLFV observables
can be cast in terms of combinations of $\mathcal{C}_{ij}^{6}$ and 
$\Lambda^{-2}$; simple, natural hypothesis on one allow to infer
constraints on the other. Table~\ref{table:effective} 
collects some bounds on the
scale of new physics (derived under the hypothesis of natural,
$\mathcal{O}(1)$, effective couplings) and on the size of the new
effective couplings (inferred for a choice $\Lambda=1$~TeV). 

{\scriptsize
\renewcommand{\arraystretch}{1.1}
\begin{center}
\begin{table}
\caption{Bounds on the effective couplings and lower bounds on the
  scale $\Lambda$ (TeV), following the hypotheses described on the text; 
the last column refers to the observable leading to the most stringent
bounds. Adapted from~\cite{Feruglio:2015}.\label{table:effective}}
\vspace*{2mm}\hspace*{3mm}
\begin{tabular}{|c|c|c|c|}
\hline
\begin{tabular}{c}
\, Effective coupling \, \\
(example)
\end{tabular}
& 
\begin{tabular}{c}
\, Bounds on ${\Lambda}$ {(TeV)} \, \\
{(for $|\mathcal{C}^{6}_{ij}| =1$)}
\end{tabular}
& 
\begin{tabular}{c}
\, Bounds on ${|\mathcal{C}^{6}_{ij}|}$ \, \\
{(for $\Lambda = 1$~TeV)}
\end{tabular}
& 
\, Observable \,  \\
\hline
$\mathcal{C}^{\mu e}_{e \gamma}$ & $6.3 \times 10^4$ &
$2.5 \times 10^{-10}$ & $\mu \to e \gamma$ \\
$\mathcal{C}^{\tau e}_{e \gamma}$ & $6.5 \times 10^2$ &
$2.4 \times 10^{-6}$ & $\tau \to e \gamma$ \\
$\mathcal{C}^{\tau \mu}_{e \gamma}$ & $6.1 \times 10^2$ &
$2.7 \times 10^{-6}$ & $\tau \to \mu \gamma$ \\
\hline
$\mathcal{C}^{\mu eee}_{\ell \ell, e e }$ & $207$ & 
$2.3 \times 10^{-5}$ & $\mu \to 3 e $ \\
$\mathcal{C}^{e \tau ee}_{\ell \ell, e e }$ & $10.4$ & 
$9.2 \times 10^{-5}$ & $\tau \to 3 e $ \\
$\mathcal{C}^{\mu \tau \mu \mu}_{\ell \ell, e e }$ & $11.3$ & 
$7.8 \times 10^{-5}$ & $\tau \to 3 \mu $ \\
\hline
$\mathcal{C}^{\mu e}_{(1,3) H\ell}$, $\mathcal{C}^{\mu e}_{He}$ & 
$160$ &
$4 \times 10^{-5}$ & $\mu \to 3 e $ \\
$\mathcal{C}^{\tau e}_{(1,3) H\ell}$, $\mathcal{C}^{\tau e}_{He}$ & 
$\approx 8$ &
$1.5 \times 10^{-2}$ & $\tau \to 3 e $ \\
$\mathcal{C}^{\tau \mu}_{(1,3) H\ell}$, $\mathcal{C}^{\tau \mu}_{He}$ & 
$\approx 9 $ &
$ \approx 10^{-2}$ & $\tau \to 3 \mu $ \\
\hline
\end{tabular}
\end{table}
\end{center}
}
\renewcommand{\arraystretch}{1.}
Despite its appeal for leading to a generic
evaluation of the new physics contributions to a given cLFV observable, 
and thus to model-independent constraints,  
there are several limitations to the effective approach. These include
taking ``natural'' values for the couplings, assuming the dominance of
a single operator when constraining a given process and the
uniqueness of the new physics scale; the latter should be kept in mind when
weighing the impact of the thus derived constraints on new physics. 

\bigskip
A second phenomenological approach consists in considering
specific new physics models or theories, and evaluating the corresponding
impact for a given class of cLFV processes. 
As extensively explored in the
literature, cLFV might be a powerful test of new physics
realisations, probing scales beyond collider reach, offering valuable
hints on properties and parameters of a given model, and allowing to
disentangle (and ultimately disfavour) between candidate models. 

A short, non-comprehensive list of examples has been presented at this
Conference; below we highlight a small subset. \\
\noindent
By studying the correlation of (high- and low-energy)
cLFV observables as predicted in the framework of generic, flavour
violating, supersymmetric (SUSY)  
extensions of the SM, one can derive useful information on the nature
of the dominant operator at
work (dipole vs. scalar); for instance, 
this is also the case of littlest Higgs models,
which can be efficiently tested via cLFV, as they lead to very
distinctive patterns for ratios of cLFV observables~\cite{Blanke:2009am}.

\begin{figure}   
\begin{center}
\begin{tabular}{cc}
\includegraphics[width=70mm]{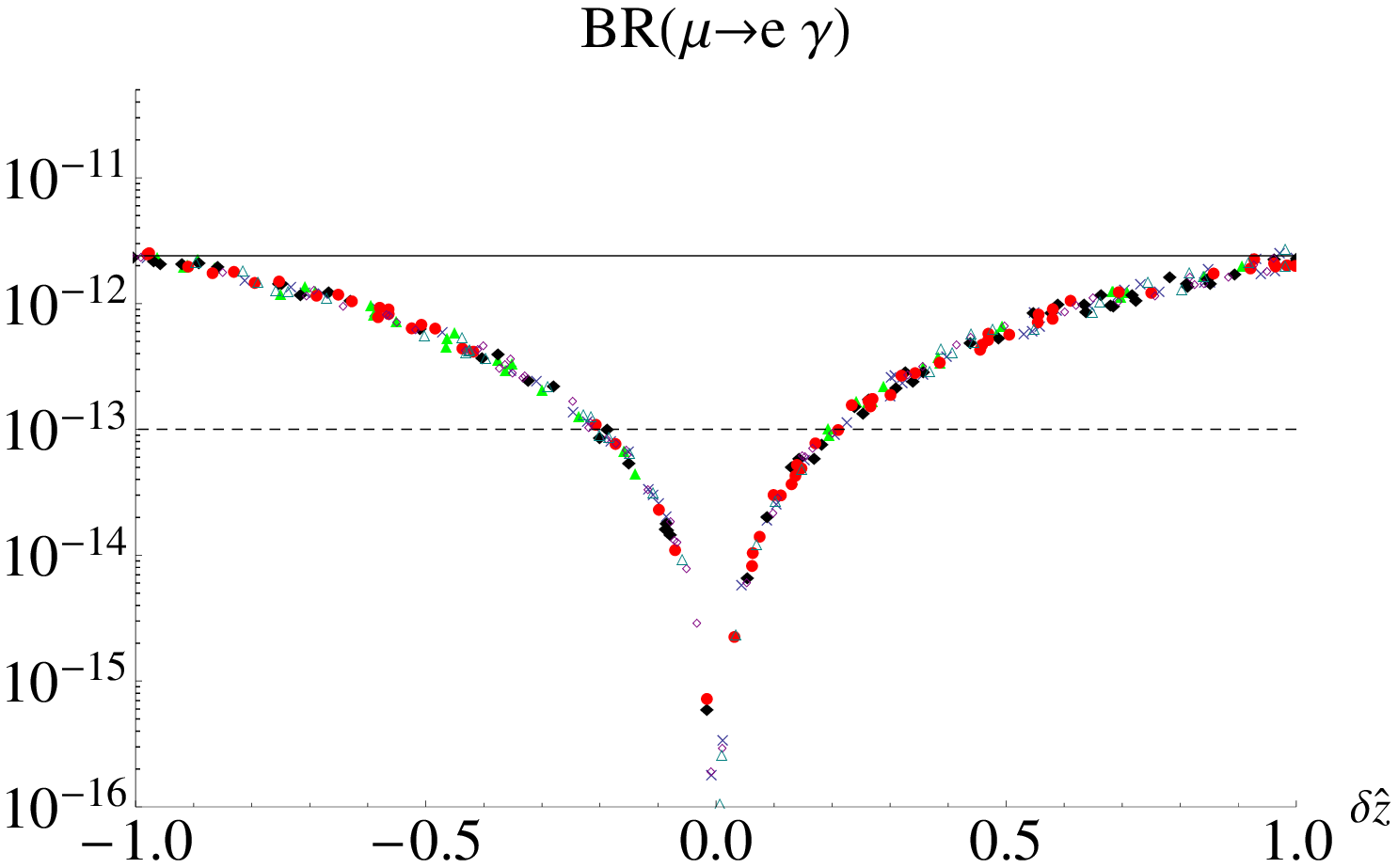}
\hspace*{4mm}&\hspace*{4mm}
\raisebox{57mm}{\includegraphics[width=60mm, angle=-90]{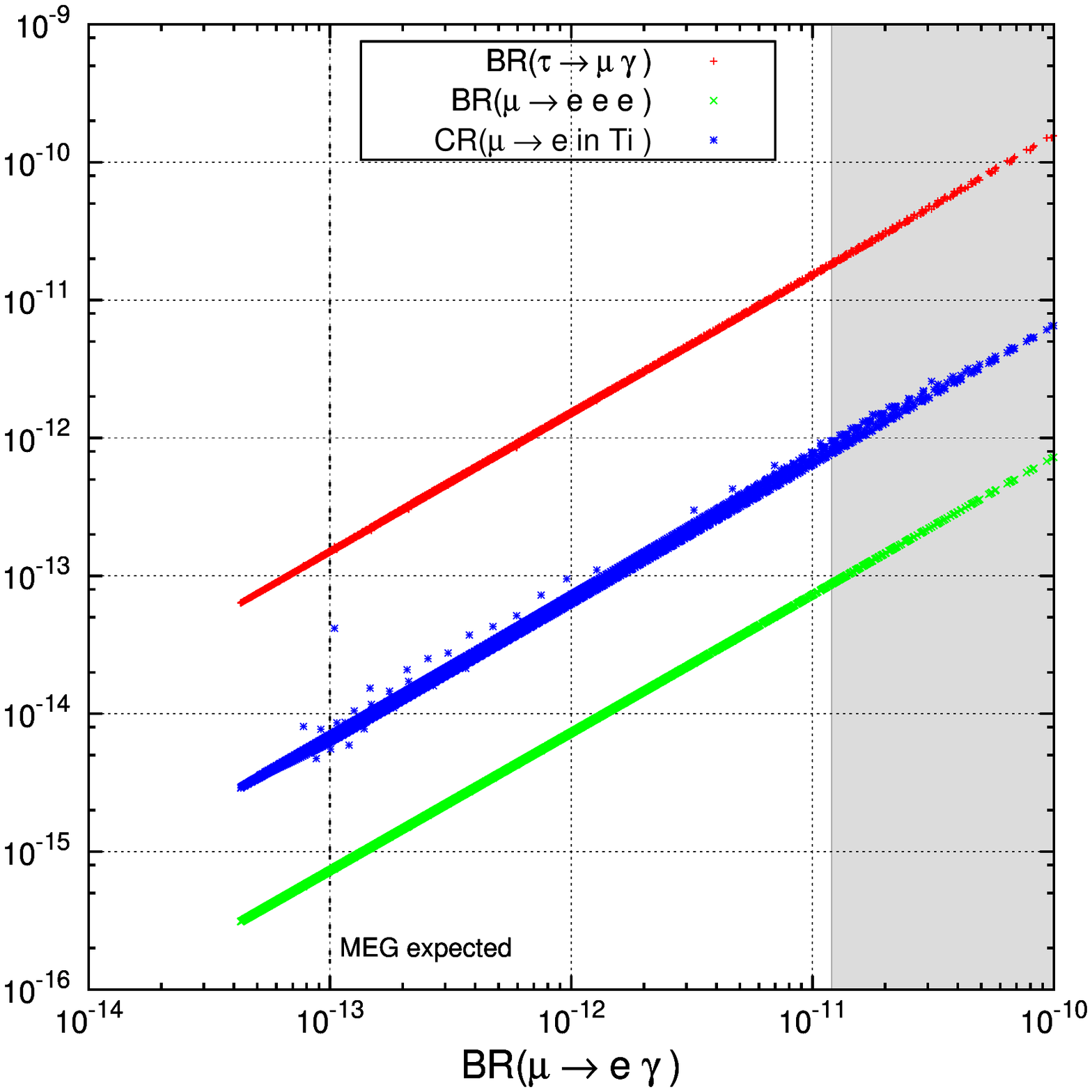}}
\end{tabular}
\end{center}
\caption{On the left, BR($\mu \to e \gamma$) as a function of the size of
 boundary kinetic terms in the framework of Holographic Composite
 Higgs models 
(from~\cite{Hagedorn.Serone}); on the
 right panel, correlation between different cLFV observables in for an  
SO(10) type II SUSY seesaw model (from~\cite{Calibbi:2009}). \label{fig:models}}
\end{figure}

In the case of specific holographic composite Higgs models, it has been
shown that although the predictions for most cLFV observables lie
below experimental reach, current bounds on BR($\mu \to e \gamma$)
allow to constrain the size of boundary kinetic terms, and thus infer
information on otherwise unreachable fundamental parameters of the
model (cf. left panel of Fig.~\ref{fig:models})~\cite{Hagedorn.Serone}.

A particularly interesting and rich example of ``geometrical cLFV''
is that of realisations of
extra-dimensional Randall-Sundrum (RS) models, with anarchic ($d=5$) Yukawa 
couplings~\cite{Beneke:2015lba}: 
current bounds from
$\mu -e$ transitions and decays already constrain the scale of new
physics to lie beyond LHC reach ($T_{\text{KK}} \gtrsim 4$~TeV,
corresponding to having the first Kaluza-Klein excitations 
$m_\text{KK}^\text{1st} \gtrsim 10$~TeV); 
future bounds should allow to exclude generic anarchic  
RS models up to 8~TeV (and correspondingly first 
excitations $m_\text{KK}^\text{1st}\gtrsim 20$~TeV). Further
examples of the constraining power of cLFV include (simple) SM extensions 
such as multi-Higgs doublet models, leptoquark constructions,
additional $Z^\prime$ bosons, etc..

Increasing the symmetry content of the model - be it in the form of a
gauge or flavour symmetry - reduces its arbitrariness, 
thus rendering the model easier to test and possibly falsify. For
instance, Left-Right symmetric models, which in addition
to exhibiting a strong interplay of cLFV and lepton number violating
observables, lead to very distinctive correlations of
observables (see, for example,~\cite{Das:2012ii}), and can thus be
easily falsifiable. 

Extended gauge groups, and in particular grand unified
theories (GUTs), in addition to possibly incorporating a mechanism of
neutrino mass generation, lead to scenarios of strong predictivity for
cLFV, as
illustrated on the right panel of
Fig.~\ref{fig:models}~\cite{Calibbi:2009}. The latter consists of a
supersymmetrisation of an SO(10) type II seesaw model - as can be
seen, data on any two observables would readily allow to exclude the
model.

\section{cLFV from seesaw realisations}
Although cLFV need not arise from the mechanism of
$\nu$ mass generation, models in which this is indeed the case - such
as the different seesaw realisations - are
particularly appealing and well-motivated frameworks. 
Whether or not a given mechanism of neutrino mass generation does have
an impact regarding cLFV stems from having non-negligble flavour violating
couplings (e.g., the Yukawa couplings) provided that the rates are
not suppressed by excessively heavy propagator masses. While
``standard'' high-scale seesaws do accommodate neutrino data with
natural values of the neutrino Yukawa couplings, the typical scale of the
mediators (close to the GUT scale) leads to a very strong suppression
of the different cLFV rates. On the other hand, 
low-scale seesaws, 
or the embedding of the seesaw in larger frameworks (as is the
case of the SUSY seesaw), are associated with a rich
phenomenology, with a strong impact regarding cLFV. 

In low-scale seesaws (as is the case of
low-scale type I seesaw, inverse and linear seesaw realisations,
...), the new ``heavy'' states do not fully decouple; their non-negligible
mixings with the light (active) neutrinos lead to the non-unitarity of the
left-handed lepton mixing matrix ($U_\text{PMNS} \to \tilde U_\text{PMNS}$),
and thus to having modified neutral and charged lepton currents. The latter
are at the origin of potentially abundant experimental/observational
signatures, which have been intensively searched for in recent years; 
negative results have
allowed to derive strong constraints on the parameter space of the new
degrees of freedom (see~\cite{Dinh:2012bp,Alonso:2012ji} for 
comprehensive discussions of cLFV in low-scale seesaws). 

A very appealing example of such low-scale models are Inverse Seesaw (ISS)
realisations: other than right-handed neutrinos, further
sterile states are added; in the case of a (3,3)
ISS realisation, three copies of each are present. 
The masses of the light active neutrinos are
given by a modified seesaw relation, $m_{\nu_i} \approx (Y_\nu v)^2
M_R^{-2} \mu_X$, where $\mu_X$ is the only source of lepton number violation 
in the model. By taking small values of $\mu_X$, one can naturally accommodate
the smallness of active neutrino masses for large Yukawa couplings and
a comparatively low seesaw scale ($M_R$ lying close to the TeV
scale). The spectrum contains, in addition
to the light states, three heavier (mostly sterile) 
pseudo-Dirac pairs, whose masses 
are given by $m_{N} \approx M_R \pm \mu_X$.
 
\begin{figure}   
\begin{center}
\begin{tabular}{cc}
\includegraphics[width=55mm,
  angle=270]{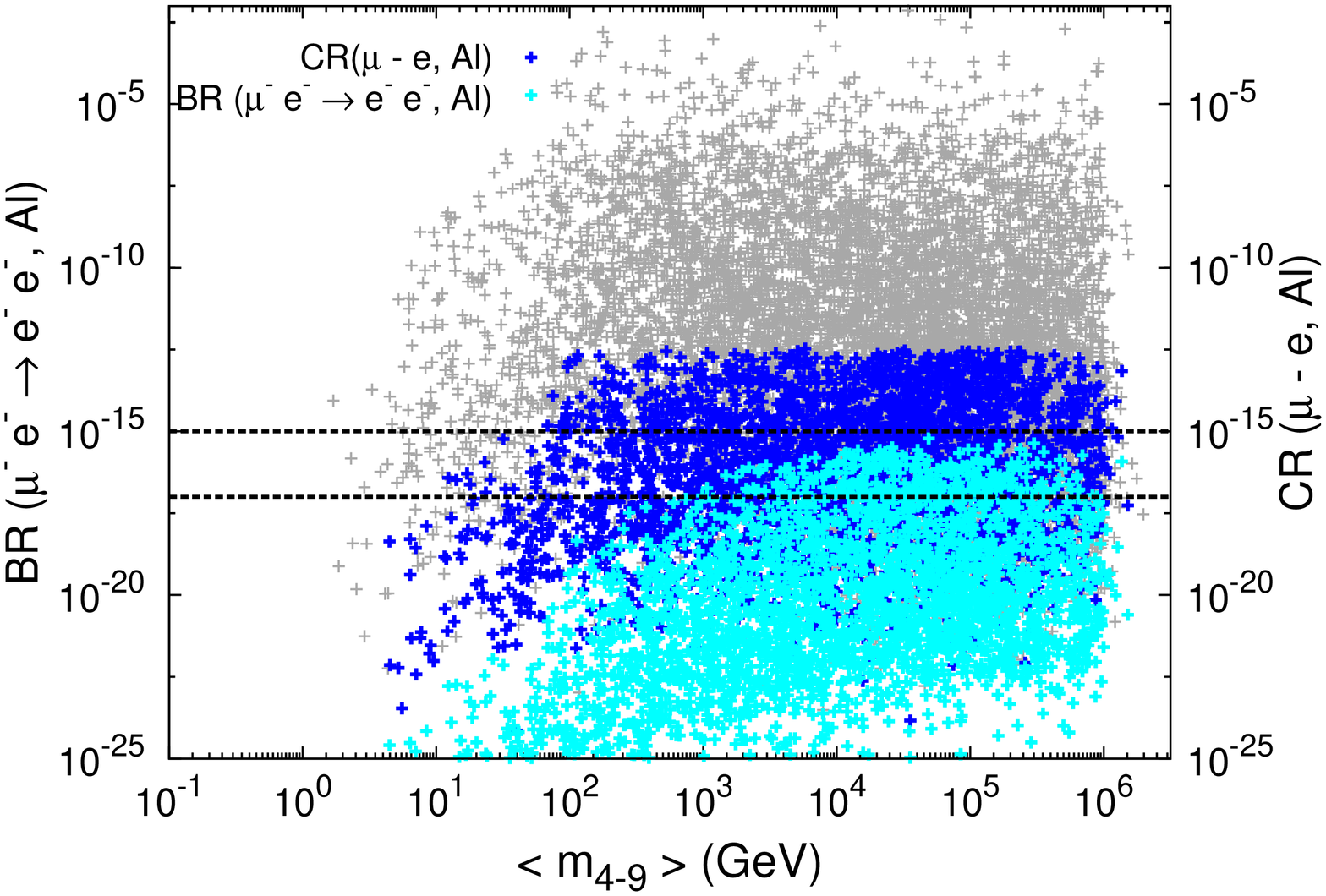} 
\hspace*{0mm}&\hspace*{0mm}
\includegraphics[width=55mm, angle=270]{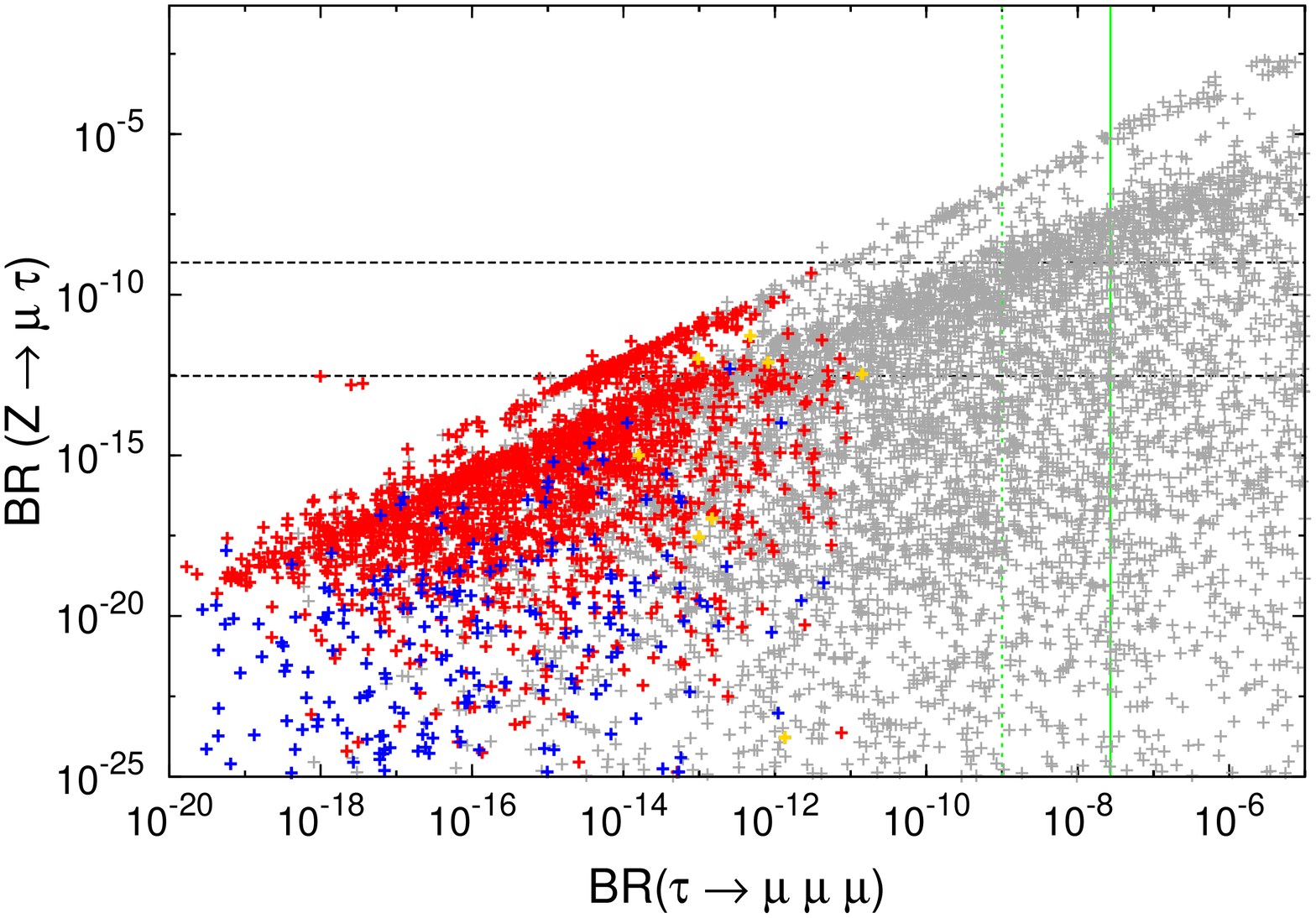}
\end{tabular}
\end{center}
\caption{
On the left panel, BR($\mu^- e^- \to e^- e^-$, Al) - cyan -  and CR($\mu-e$, Al)
- blue - as a function of the average mass of the heavier, mostly sterile
states, in a (3,3) ISS realisation. Horizontal lines denote future
experimental sensitivities (from~\cite{Abada:2015oba}).
On the right, BR($Z \to \tau \mu$) vs. BR($\tau \to 3 \mu$) in a (3,3)
ISS realisation. Vertical lines denote future experimental
sensitivities while the horizontal ones correspond to the prospects of
a GigaZ facility and of the FCC-ee (from~\cite{Abada:2014cca}).
In both cases, grey points are phenomenologically excluded.
\label{fig:iss:low}}
\end{figure}

The (3,3) ISS opens the door to a very rich phenomenology, which
includes abundant cLFV signatures, both at low- and at 
high-energies (see, for
example,~\cite{Abada:2013aba,Abada:2014cca,Abada:2015oba}).  
To illustrate the potential impact regarding high-intensity
facilities, the left panel of Fig.~\ref{fig:iss:low} displays the
prospects for $\mu - e$ conversion, as well as the Coulomb enhanced decay of
a muonic atom (both for the case of Aluminium targets), as a function
of the average mass of the heavier states, $<m_{4-9}>$. Although
CR($\mu - e$, Al) is in general associated to
larger rates, for sterile
states above the TeV, both observables are expected to be well within
reach of the COMET experience (horizontal lines respectively denoting
the sensitivity of Phase I and II), or of the Mu2e experiment. 

At higher energies (for example, in the case of a future circular
collider, as FCC-ee), one can also explore cLFV in the decay of
heavier states, as for instance in $Z \to \ell_i \ell_j$. 
In the ISS (3,3) realisation, especially in
the ``large'' sterile mass regime, the cLFV $Z$ decays exhibit a strong
correlation with cLFV 3-body decays (since the latter 
are dominated by the $Z$-penguin
contribution). The prospects for a (3,3) ISS realisation, for the
case of $\mu -\tau$ flavour violation, are
shown in the right panel of Fig.~\ref{fig:iss:low}. Not only
can one expect to have BR($Z \to \tau \mu$) within FCC-ee reach,
but this observable does allow to probe $\mu -\tau$ flavour violation
well beyond the sensitivity of a future SuperB factory (large values
of BR($\tau \to 3 \mu$) are precluded in this realisation due to the
violation of other cLFV bounds).  

At the LHC, searches for heavy ISS mediators relying on cLFV
signatures can be carried; as recently proposed, a significant number
of events (after cuts) could be expected from the channel 
$q q^\prime \to \tau \mu +2$jets (no missing $E^T$)~\cite{Arganda:2015ija}. 

\begin{figure}   
\begin{center}
\hspace*{-5mm}
\begin{tabular}{cc}
\includegraphics[width=83mm]{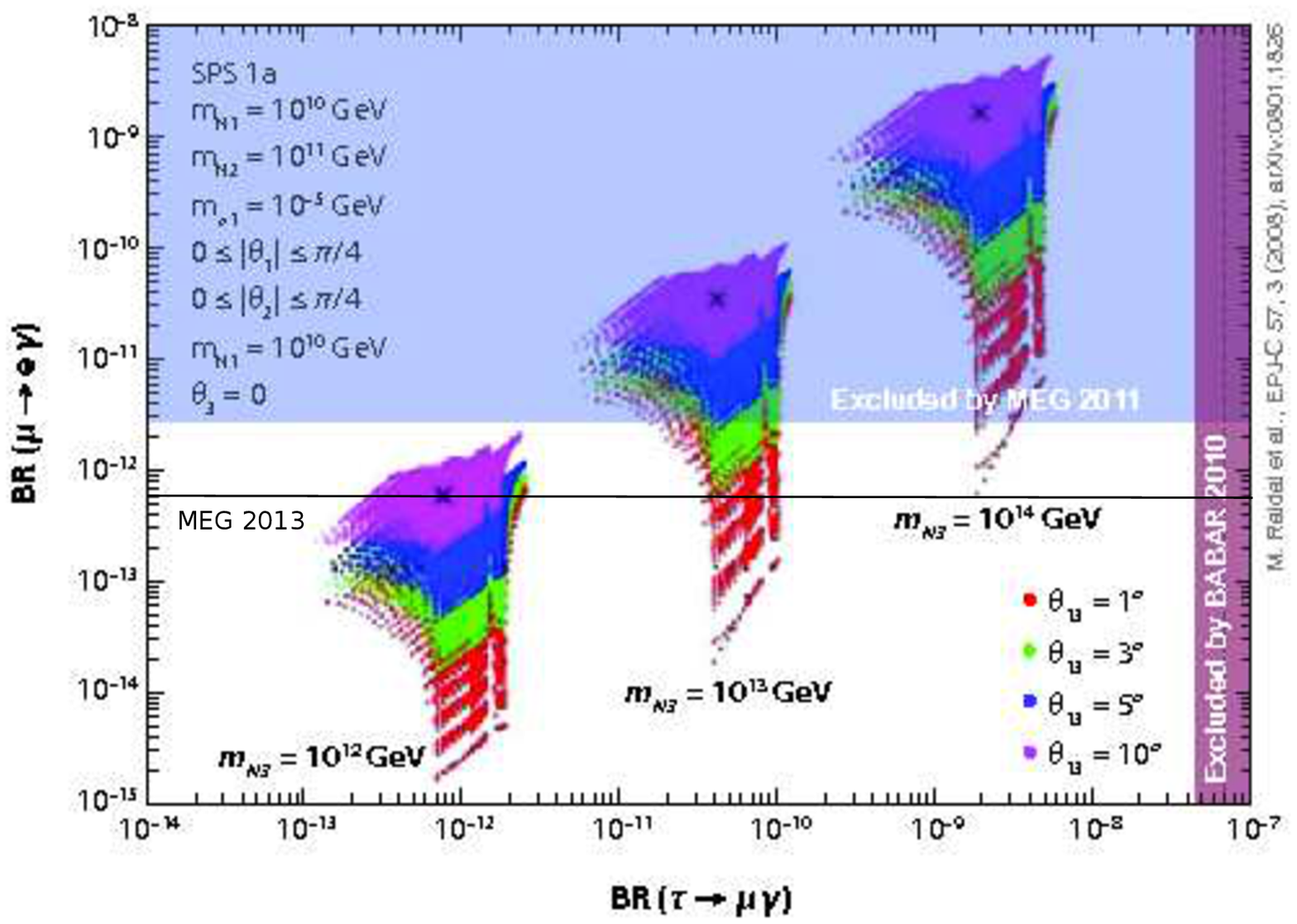}
\hspace*{-2mm}&\hspace*{-3mm}
\raisebox{-4mm}{\includegraphics[width=83mm]{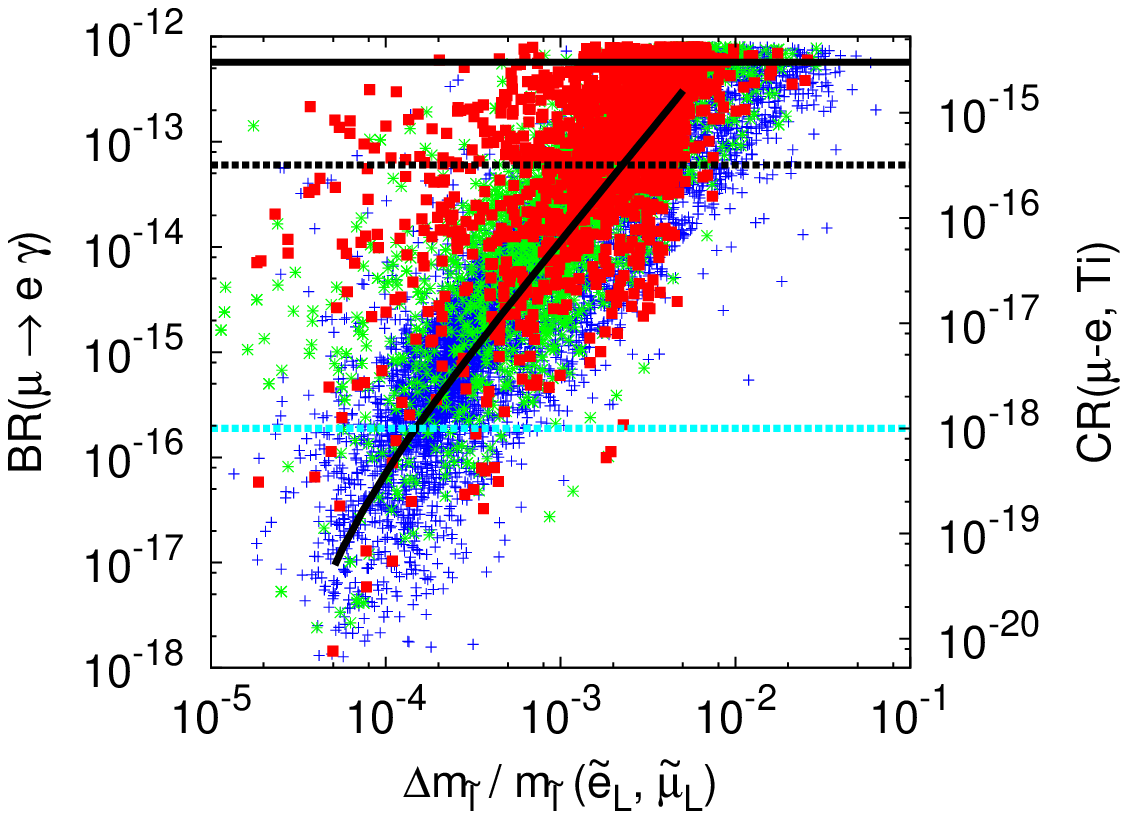}}
\end{tabular}
\end{center}
\caption{
On the left, correlation between BR($\mu \to e \gamma$) and BR($\tau
\to \mu \gamma$) in a type I SUSY seesaw for different 
seesaw scales (from~\cite{Antusch:2006vw});
on the right panel, $1^\text{st}$ and $2^\text{nd}$ generation 
charged slepton mass splittings vs. BR($\mu \to e \gamma$), with CR($\mu -e$,
Ti) on secondary y-axis in a type I SUSY seesaw, for different values
of the heaviest 
right-handed neutrino mass $M_{R_3} = 10^{13,14,15}$~GeV ($M_{R_1,R_2} =
10^{10,11}$~GeV) and for a flavour conserving modified mSUGRA benchmark 
(from~\cite{Figueiredo:2013tea}).
\label{fig:susyseesaw}}
\end{figure}

\bigskip
Another rich and well-motivated framework leading to observable cLFV
is that of the SUSY seesaw (a high-scale seesaw embedded in the
context of otherwise flavour conserving SUSY models). 
In the case of a type I SUSY seesaw~\cite{Borzumati:1986qx}, sizeable 
neutrino Yukawa
couplings (as characteristic of a high-scale seesaw) and the
possibility of new, not excessively heavy mediators (the SUSY partners),
open the door to large contributions to cLFV observables. Having a
unique source of flavour violation implies that the observables
exhibit a high degree of correlation; such a synergy can be explored,
allowing to put the seesaw hypothesis to the test and possibly hinting on
certain parameters. The complementarity of two low-energy observables
is depicted on the left panel of Fig.~\ref{fig:susyseesaw}: 
BR($\mu \to e \gamma$) vs. BR($\tau \to \mu \gamma$), for different
seesaw scales (and for distinct values of the then unknown Chooz
angle)~\cite{Antusch:2006vw}. 
The determination of these two observables, in association with
the discovery of SUSY, would allow to infer
information on the seesaw scale $M_R$, 
or then readily disfavour the SUSY seesaw as the
source of cLFV. The potential of exploring the interplay of
high-intensity (for instance $\mu \to e \gamma$ and  
$\mu-e$ conversion) and collider
observables (for example, the splittings between the first and second
generation charged slepton masses, $\Delta m_{\tilde \ell}$) 
is summarised on the right
panel of Fig.~\ref{fig:susyseesaw}: ``isolated'' cLFV manifestations
(i.e., outside the coloured regions) would allow to disfavour the SUSY
seesaw hypothesis as the (unique) underlying source of lepton flavour
violation, while ``compatible'' ones would strenghten it, 
furthermore hinting on the seesaw scale~\cite{Figueiredo:2013tea}. 

\section{Outlook}
As of today, we have firm evidence that flavour is violated in
the quark sector, as well as in the neutral lepton one. In the absence
of a fundamental principle preventing it, there is no
apparent reason for Nature to conserve charged lepton flavours. By
itself, any observation of a cLFV process would constitute a clear
signal of new physics - beyond the SM extended via massive (Dirac)
neutrinos. 
As we aimed at illustrating in the present brief review, cLFV
observables could provide valuable (indirect) information on the
underlying new physics model, and certainly contribute to at least
disfavour several realisations. 

The current (and planned) experimental programme, with numerous
observables being searched for in a large array of high-intensity and
high-energy experiments (see~\cite{Ootani.Neutrino16}) 
renders cLFV a privileged laboratory to search for new physics.

\ack
AMT is gratefull to the Organisers of ``Neutrino 2016''
for the invitation and support.
Part of the work here summarised was done within the framework of the
European Union's Horizon 2020 research and innovation programme under
the Marie Sklodowska-Curie grant agreements No 690575 and No 674896.

\end{document}